\documentclass[accepted]{uai2026} 
                        
\usepackage[american]{babel}

\usepackage{natbib} 
\usepackage{mathtools} 
\usepackage{amssymb}
\usepackage{multirow}
\usepackage{xcolor}
\usepackage{booktabs} 
\usepackage{tikz} 
\usepackage{comment}
\usepackage{makecell}
\usepackage{subcaption}

\title{A Causal Framework for Evaluating ICU Discharge Strategies}

\author[1]{\href{mailto:<s.n.simha@amsterdamumc.nl>?Subject=Your UAI 2026 paper}{Sagar Nagaraj Simha}{}}
\author[1]{Juliette Ortholand}
\author[2]{Dave Dongelmans}
\author[3,4]{Jessica D. Workum}
\author[6]{Olivier W.M. Thijssens}
\author[1]{Ameen Abu-Hanna}
\author[1,5]{Giovanni Cinà}
\affil[1]{%
    Department of Medical Informatics, Amsterdam UMC, University of Amsterdam, NL
}
\affil[2]{%
    Department of Intensive Care Medicine, Amsterdam UMC, University of Amsterdam, NL
}
\affil[3]{%
    Department of Intensive Care, Elisabeth-TweeSteden Hospital, Tilburg, NL
  }
\affil[4]{%
    Department of Intensive Care, Erasmus MC, Rotterdam, NL
  }
\affil[5]{%
Institute of Logic, Language and Computation, University of Amsterdam, NL
}
\affil[6]{%
Pacmed, Amsterdam, NL
}

\begin{document}
\maketitle
\begin{abstract}
 In this applied paper, we address the difficult open problem of when to discharge patients from the Intensive Care Unit. This can be conceived as an optimal stopping scenario with three added challenges: 1) the evaluation of a stopping strategy from observational data is itself a complex causal inference problem, 2) the composite objective is to minimize the length of intervention and maximize the outcome, but the two cannot be collapsed to a single dimension, and 3) the recording of variables stops when the intervention is discontinued.
 Our contributions are two-fold. First, we generalize the implementation of the g-formula Python package, providing a framework to evaluate stopping strategies for problems with the aforementioned structure, including positivity and coverage checks. Second, with a fully open-source pipeline, we apply this approach to MIMIC-IV, a public ICU dataset, demonstrating the potential for strategies that improve upon current care.
\end{abstract}

\section{Introduction}\label{sec:intro}

Determining when to stop an ongoing intervention is a fundamental problem in a variety of fields. Examples are the optimal duration of a therapy in healthcare \citep{daneman2016duration}, the secretary problem in mathematical economics \citep{ferguson1989solved}, monitoring/maintenance problems in operations research \citep{rust1987optimal}, among others. 
Stopping strategies arise as a special class of dynamic treatment regimes (DTRs) \citep{chakraborty2013statistical}
, where the decision variable is not the choice of intervention but whether to discontinue. Unlike static or fixed-duration strategies, dynamic stopping strategies depend on the temporal evolution of continuously monitored covariates, increasing the complexity of their evaluation. 


The Intensive Care Unit (ICU), the hospital ward were the most severe patients are treated, requires clinicians to continuously take optimal stopping decisions. 
The decision of when to stop ICU care--and send a patient to a step-down ward-- is one of them, and can be conceived as an optimal stopping problem. While a longer ICU stay may benefit survival, an extra ICU day is very expensive and, in case of congestion of ICU beds, may prevent the admission of another patient \citep{long2018boarding}. Optimising ICU discharge requires balancing length of stay and probability of survival. 
Current guidelines on ICU discharge outline general best practices; however, they do not specify a concrete operational protocol and are supported by limited empirical evidence \citep{nates2016icu}. A recent survey including ICU clinicians from 40 countries found out that only about half of the ICUs had discharge protocols in place \citep{hiller2024current}, and a survey on Reinforcement Learning applications to the ICU highlighted the evaluation procedure as one of the weak points in this body of literature \citep{otten2024does}. This motivates the need for a methodologically robust way to evaluate candidate discharge strategies.

\begin{figure}[t!]
    \centering
    \includegraphics[width=\columnwidth]{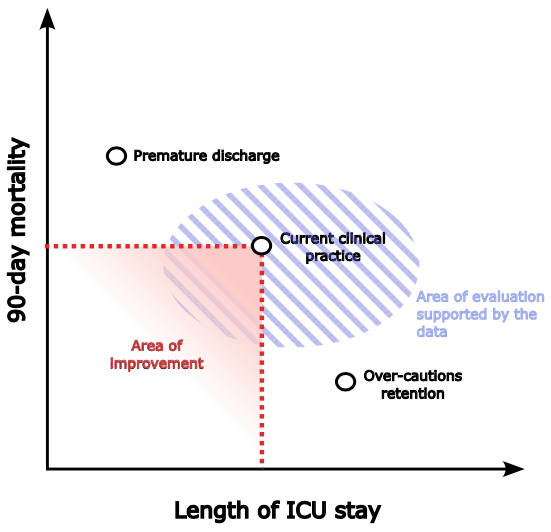}
    \caption{The ICU discharge trade-off represented in the space of potential outcomes for our multi-objective stopping problem. Given a fixed dataset, each strategy returns an average 90-day mortality on the y-axis as well as an average length of ICU stay on the x-axis. The red box defines the set of strategies that constitute an improvement over current clinical practice, i.e. strategies leading to lower mortality and lower utilization of ICU beds. The blue dashed area defines the set of strategies we are actually able to evaluate given the available data.
    }
    \label{fig:eye-catcher}
\end{figure}

Figure \ref{fig:eye-catcher} can be used as an aid to conceptualize our work. Given a fixed dataset, each strategy's potential outcome is a point in the 2-dimensional space spanning the two outcomes of interest: 90-day mortality and length of ICU stay (both averaged across the data). We are interested in finding strategies constituting an improvement over current care, i.e. in the red area. However, our data may not allow for the evaluation of every strategy.

The evaluation of ICU discharge strategies is complicated by (at least) three additional challenges. First, given the difficulty of testing new potential strategies directly on patients, the evaluation of strategies often requires causal inference approaches to estimate effects from observational data. Second, when the outcome of interest is e.g. mortality, it is often not straightforward to collapse the cost of the outcome and the cost of continuing the intervention to a single dimension of `utility'. Finally, since the ICU entails a high level of monitoring, stopping the intervention and discharging means terminating the measurement of certain variables.


\paragraph{Contributions.}
When it comes to handling these challenges, many existing computational approaches are limited (see Related Works below). In this applied paper, we address these limitations as follows
: (i) we formulate a causal estimand for the comparative evaluation of dynamic, covariate-dependent stopping strategies; (ii) we implement this estimand by extending the \texttt{pygformula} \citep{mcgrathGfoRmulaPackageEstimating2020} package to evaluate optimal stopping rules when variables are censored at stopping; (iii) we provide a series of diagnostics to understand whether the strategy of interest can be evaluated on the available data, and (iv) we apply the framework in a target trial emulation evaluating hypothetical ICU discharge strategies using a public ICU dataset, the MIMIC-IV database \citep{PhysioNet-mimiciv-3.1}.


In our application, ICU discharge decisions are modelled as a sequential stopping process in which, at each decision time, a strategy $g$ determines whether ICU care is continued or terminated based on patient history. Discharge is treated as a censoring intervention: once the stopping criterion is met, the patient exits the ICU, variables are not measured any more and post-discharge outcomes are evaluated at each time point. For each candidate discharge strategy, assuming all assumptions are met and a dataset is fixed, our implementation returns both an average outcome and an average length of the intervention, as well as an assessment on whether the data supports the evaluation. All our code is available open-source\footnote{Our code is available in the Supplementary files.}.


\section{Related Work}


Evaluating clinical guidelines to optimise outcomes following ICU discharge 
\citep{nates2016icu, knightNurseledDischargeHigh2003}, requires taking into account time-varying confounding while avoiding conditioning on post-treatment variables, using a causal inference approach. Unfortunately, when attempting to optimise ICU decisions, such causal methods are not always fully adhered to  \citep{komorowskiArtificialIntelligenceClinician2018a, lejarzaOptimalDischargePatients2023, thoralExplainableMachineLearning2021, otten2024does}. For this reason, guidelines for the use of Reinforcement Learning in ICU have raised awareness regarding data availability (positivity assumption) and variables measured (no unmeasured confounders assumption) \citep{gottesmanGuidelinesReinforcementLearning2019}. 
Mindful of this warning, rather than searching for an optimal strategy as Causal Reinforcement Learning generally aims to do \citep{cliftonQLearningTheoryApplications2020}, we aim in this paper to properly evaluate candidate strategies against current care.  

To contrast strategies for treatment stopping, several causal approaches have been used: inverse probability weighting \citep{keoghCausalInferenceSurvival2023a} that reweights the trajectories, g-estimation \citep{moodieEstimatingResponseMaximizedDecision2009a} that directly model treatment effect and g-computation with g-formula that focus on modelling the conditional mean outcome \citep{robinsNewApproachCausal1986, robinsErrataNewApproach1987, taubmanInterveningRiskFactors2009, youngComparativeEffectivenessDynamic2011, smithEmulationTargetTrial2022, wanisEmulatingTargetTrials2023}. Analysis are then designed using emulation of target trial on observational data \citep{hansfordReportingObservationalStudies2023, smithEmulationTargetTrial2022, garcia-albenizImmediateDeferredInitiation2015a}. 

G-computation offer the most natural way to simulate various strategies, and has been implemented on open source packages such as \texttt{pygformula} \citep{mcgrathGfoRmulaPackageEstimating2020}. More recently, the G-computation approach has been extended to more flexible modelling with deep learning approaches \citep{xiongGTransformerCounterfactualOutcome2024, suCounterfactualSepsisOutcome2024, dengUncertaintyQuantificationConditional2024}. Yet, modelling structured data using deep learning approaches has been shown to offer a small improvement at the cost of interpretability, need for data and, computation time \citep{varoquauxHypeSustainabilityPrice2025, maheuxForecastingIndividualProgression2023, christodoulouSystematicReviewShows2019a}. 

To our knowledge, none of these approaches have considered different models for the outcome with and without treatment, which offer model flexibility and is arguably a requirement to evaluate ICU discharge. Indeed, deaths in and outside of the ICU are different outcomes and many vital signs monitored in ICU are not recorded outside of it (or recorded at a different granularity) due to the lower level of monitoring.

\section{Methods}
We introduce here the notation, the mathematical framework and the experimental design. We further use the DTR terminology as described in \cite{deliuDynamicTreatmentRegimes2022}.

\subsection{Notation}

We employ a discrete representation of time $t = 0,1,\dots,T$, with $T$ the end of follow-up time. Each patient $i$ is associated with the outcome of interest (90-day all-cause mortality in our case), a binary variable noted $y_{i,t} \in Y_t$ such that $y_{i,t} = 1 \;\Rightarrow\; y_{i,t'} = 1 \text{ } \text{for all } t' > t$. We define the time when the outcome is determined as $t_d = \text{argmin}\{t,\text{ if } \exists t, y_{i,t} = 1\text{ or $T$ if } \forall t,\, y_{i,t} = 0 \}$, using survival analysis formalism.
A set of baseline and longitudinal covariates, $\ell_{i,t}\in L_t$, is also associated with each patient $i$ at each decision time $t$. In our setup the covariates are only defined if $y_{i,t} = 0$.

If $y_{i,t} = 0$, we consider a binary treatment $a_{i,t} \in \{0,1\}$ indicating whether treatment is \emph{continued} ($1$) or \emph{stopped} ($0$) at time $t$ for patient $i$ and denote with $\bar{a}_{i,t} = (a_{i,0}, ..., a_{i,t})\in \bar{A}_t$ the choices made since baseline up to time $t$. A treatment strategy, $g = (g_0,\dots,g_T)$, is associated with it such that:
\begin{align*}
    &a_{i,0}^{g} = g_0(\ell^g_{i,0}),  a_{i,t}^{g} = g_t(\bar{\ell}^g_{i,t}, \bar{a}^g_{i,t-1}), 
\text{ } t = 1,\dots,T \\
&a_{i,t}^{g} = 0 \;\Rightarrow\; a_{i,t'}^{g} = 0 \text{ } \text{for all } t' > t
\end{align*}
assuming that once treatment stops, it remains stopped.
We will further refer to such treatment strategy as a DTR following \cite{murphyMarginalMeanModels2001} definition. Static strategies are a subset of dynamic strategies incorporating only baseline covariates.  Following standard terminology we refer to the strategy observed in the data as the ``natural course'' strategy and denote it with $g_N$. We consider a finite-horizon dynamic treatment strategy ending at time $T$. We denote with $\tau_{i}^{g} = \text{argmax}_t\{t \text{ such that }a^g_{i,t}=1\}$ the length of treatment for patient $i$; this value can at max be $T-1$ since at $T$ we only evaluate the outcome. 

\subsection{Estimand (task)}

Using the potential outcome notation, we are looking for the set of DTRs $g$ that maximise $E\left[Y^{g_N}_{T}(\bar{A}^{g_N}_{T-1}) - Y^{g}_{T}(\bar{A}^{g}_{T-1})\right]$ while minimising $E[\tau^g - \tau^{g_N}]$ across the population. That is, the objective is to maximize improvement in the outcome with respect to natural course, while also reducing length of treatment. Such strategies have a potential outcome within the red area in Figure \ref{fig:eye-catcher}. 

\subsection{Assumptions}\label{sect:assumption}

Standard causal inference assumptions are stable unit treatment value, no unmeasured confounders and positivity  \citep{deliuDynamicTreatmentRegimes2022} (see \ref{sect:appendix_assumptions} for further details). If the two first assumptions come from the study design and can hardly be tested, positivity is a major issue that should be assessed from the data. 

Classic approaches define the set of history-compatible trajectories at
epoch $t$, $\mathcal{D}^g_t$; a subgroup of those are trajectories for which the decision at time $t$ coincide with strategy $g$, we call such subset $\mathcal{M}^g_t$. Metrics build on these sets can be defined such as the \emph{Match Count} $|\mathcal{M}^g_t|$ or the \emph{Coverage rate}, the fraction of observations at time $t$ for which the clinician's action coincides with the target strategy's prescription $\rho^g_t = \frac{|\mathcal{M}^g_t|}{|{\mathcal{D}^g_t}|}$. Values of $\rho^g_t$ close to one indicate that the target strategy largely agrees with observed clinical practice at that epoch, a necessary (though not sufficient) condition for positivity. It should be read alongside with the match count because the apparent agreement may be an artifact of selection rather than genuine overlap.

The metrics above are summed over trajectories, thus might fail to spot an under-represented type of patient. As dynamic strategies are triggered by patient characteristics, this might pose a problem if the set of patient triggering a strategy is not well-represented. If for example a discharge strategy is overly conservative compared to natural course, all patient will be discharged before they can trigger the conservative strategy, and we won't be able to really evaluate such strategy. To uncover such cases, we have computed the above metrics separately depending on treatment decision at time $t$ for dynamic strategies. In Appendix \ref{sect: appendix_positivity_sn} we perform additional checks e.g. on Effective Sample Size ratio and visualisation of positivity violations. 

\subsection{Modelling}

Longitudinal observational data may contain treatment-confounder feedback, i.e, when past treatment affects subsequent confounders, which then influence later treatment decisions.
The Parametric G-formula is a causal inference method designed to tackle this problem, allowing for the estimation of the effects of time-varying treatment strategies using such observational datasets. The method is based on Robins’ g-formula \citep{robinsNewApproachCausal1986, robinsErrataNewApproach1987}, which characterizes the distribution of outcomes under a specified treatment strategy in terms of the full set of conditional densities of the observed data. 
Under the identification assumptions mentioned above, we identify $E[Y_{t_d}(\bar{A}^{g}_{t_d-1})]$ by 
\begin{align*}
 \psi(g) =\sum_{\bar{\ell}_{t_d-1}\in \bar{L}} \sum_{\bar{a}_{t_d-1}\in \bar{A}}\
&E\!\Big[Y^g_{t_d}\mid
\bar{\ell}^g_{t_d-1},\,\bar{a}^g_{t_d-1},\bar{y}^g_{t_d-1} = \bar{0}\Big] \\
 &\times f(\bar{\ell}^g_{t_d-1},\bar{a}^g_{t_d-1}, \bar{y}^g_{t_d-1}= \bar{0})
\end{align*}
where $\bar{0}$ is short-hand for a history of survival up to that point, and $f$ the probability density.

Let us shorten said density as $f(...)$. In the classic set-up, this joint distribution can be factorised  assuming that at each draw, death is first sampled, then in case of survival covariates are sampled next and finally the treatment is chosen with a deterministic dynamic treatment strategy $g$, yet all that, almost instantaneously.
Yet in our setup, due to the difference in terms of outcome occurrence when under or not under treatment, we posit two outcome models: one for in-ICU death that has access to all the covariates and one for post-ICU death that only has access to covariates observed until discharge at $\tau^g$. We assume that conditioning on covariates observed during treatment and current treatment history is enough to get the independence at further steps. We can then rewrite the joint distribution as: 
\begin{align*}
f(...)
&=\Big[\prod_{t=0}^{\tau_g}
f\bigl(
\ell_t^g \mid \bar{\ell}^g_{t-1},
\bar{a}^g_{t} = \bar{1},\bar{y}^g_{t} = \bar{0} 
\bigr)\Big]\\
&\times \Big[\prod_{t=0}^{\tau_g}
f\bigl(
\bar{y}^g_{t} = \bar{0} \mid \bar{\ell}^g_{t-1},
\bar{a}^g_{t-1} = \bar{1},\bar{y}^g_{t-1} = \bar{0}
\bigr)\Big]\\
&\times\Big[\prod_{t=\tau^g+1}^{t_d}
f\bigl(
\bar{y}^g_{t} = \bar{0} \mid \bar{\ell}^g_{\tau^g},
\bar{a}^g_{t-1},\bar{y}^g_{t-1} = \bar{0}
\bigr)\Big]
\end{align*}
As often done in g-formula computation, we assume a stable Markov process, i.e. we take the causal mechanisms to be stable across time points. We specify a model for each $f\bigl(
L^g \mid \bar{\ell}^g,
\bar{a}^g = \bar{1},\bar{y}^g = \bar{0} 
\bigr)$, $f\bigl(Y^g \mid \bar{\ell}^g, \bar{a}^g = \bar{1}, \bar{y}^g= \bar{0}\bigr)$ and $f\bigl(Y^g \mid \bar{\ell}^g_{\tau_i^g}, \bar{a}, \bar{y}^g= \bar{0} \bigr)$ , using a generalised linear model estimated with the package \texttt{statsmodels} \citep{seabold2010}. Distributional families are further specified in Appendix \ref{sect:history} along with the summarisation of the history.


\subsection{Non-Iterative Conditional Expectation Estimator (NICE)}

We use the Non-Iterative Conditional Expectation (NICE) estimator \citep{robinsNewApproachCausal1986, robinsErrataNewApproach1987} as implemented in the \texttt{pygformula} package \citep{mcgrathGfoRmulaPackageEstimating2020}. It is composed of three steps: (i) estimating the density models from the observed data, (ii) simulating patients trajectory under strategy $g$ with forward Monte Carlo sampling, (iii) computing $\hat{\psi}(g)$ using the simulated final outcomes. 
The current implementation does not allow for different models for the outcome with and without treatment. To accommodate that, we updated the simulation algorithm.


    
\subsubsection{Monte Carlo Simulation}

The Monte Carlo simulation under the intervention strategy $g$ takes place for each subject $i = 1,\dots,N$ over $t = 0,\dots,T$ and noted $\ell_{i,t}^{g,*}$:
\paragraph{Initialization ($t = 0$):} draw $\ell_{0,i}^{*}$ from the empirical distribution of $L_0$ in the observed data and evaluate $a_{0,i}^{g,*}$.
 \paragraph{Recurrence ($T\geq t \geq 1$):} 
 
 \begin{itemize}
    \item Sample the outcome:
    \begin{itemize}
        \item if $a_{i,t-1}^{g,*} = 1$:
        \begin{align*}
        y_{i,t}^{g,*} \sim f\bigl(Y^g \mid \bar{\ell}^{g,*}_{t-1}, \bar{a}^{g,*}_{t-1} = \bar{1}, \bar{y}^{g,*}_{t-1}= \bar{0}\bigr)
        \end{align*}
        \item if $a_{i,t-1}^{g,*} = 0$:
        \begin{align*}
        y_{i,t}^{g,*} \sim f\bigl(Y^g \mid \bar{\ell}^g_{\tau_i^{g,*}}, \bar{a}^{g,*}_{t-1}, \bar{y}^{g,*}_{t-1}= \bar{0} \bigr)
        \end{align*}
    \end{itemize}
    \item If $y_{i,t}^{g,*} = 1$: 
    \begin{itemize}
        \item set $y_{i,T}^{g,*} = 1$ and stop the simulation,
    \end{itemize}
    \item If $y_{i,t}^{g,*} = 0$:
    \begin{itemize}
        \item If $a_{i,t-1}^{g,*} = 1$: sample the covariates 
    \begin{align*}
        \ell_{i,t}^{g,*}
    \sim
    f\bigl(
L^g \mid \bar{\ell}^{g,*}_{t-1},
\bar{a}^{g,*}_{t} = \bar{1},\bar{y}^{g,*}_{t} = \bar{0} 
\bigr)
    \end{align*}
    and determine the new treatment:
    \begin{align*}
    a_{i,t}^{g,*} = g_t\bigl(\bar{\ell}_{i,t}^{g,*}, \bar{a}_{i,t-1}^{g,*}\bigr).
    \end{align*}
    \item Incrementation:
    \begin{itemize}
        \item If $t = T$ : stop the simulation,
    \item If $t < T$ increment $t \leftarrow t+1$.
    \end{itemize}
    
    \end{itemize}
 \end{itemize}
 
\subsubsection{Mean Outcome Estimator}

The causal mean outcome of 90-day mortality is approximated via simulation over the $N$ synthetic subjects:
\begin{align*}
    \hat{\psi}(g)
=
\frac{1}{N}
\sum_{i=1}^{N}
y_{i,T}^{g,*} =
\frac{1}{N}
\sum_{i=1}^{N}
y_{i,t_d}^{g,*}
\end{align*}
The average length of stay in ICU is obtained similarly via the simulation.

\subsection{Experimental Design}

\subsubsection{Data and Pre-processing: MIMIC-IV}
Data for this study were derived from the Medical Information Mart for Intensive Care (MIMIC-IV, v4.1) database \citep{PhysioNet-mimiciv-3.1} that comprises adult ICU admissions recorded at Beth Israel Deaconess Medical Center (Boston, MA) between 2008 and 2019. 
We used the BlendedICU pipeline \citep{oliverIntroducingBlendedICUDataset2023a} to further standardize patient-level variables and ensure consistent representation of demographics, drug exposures, laboratory measurements, and physiological time-series data. Each admission preserves its full, variable-length ICU trajectory, allowing patient-specific follow-up durations.

\subsubsection{Target Trial Emulation} \label{sect:tte}

We frame the evaluation of ICU discharge strategies as a target trial emulation, in which candidate strategies are evaluated against the clinical practice implicitly observed in the data. The target trial is defined by the following components \citep{cashinTransparentReportingObservational2025}.

\paragraph{Inclusion and exclusion criteria}
We exclude stays shorter than 12 hours and stays which have baseline characteristics missing - age, sex, and admission origin.


\paragraph{Outcome of interest} The binary outcome of interest $Y$ is 90-day all-cause mortality defined as death occurring either inside the ICU or after discharge within follow-up time $T =$ 90 days \citep{schoenfeldSurvivalMethodsIncluding2005}.  We estimate the total effect of ICU discharge strategy from 12 hours after ICU admission thus we thus only study patients alive after 12 hours. The associated causal question  can be formulated as \textit{If the patient survives the first 12 hours, what is the average effect of a discharge strategy on 90 days all-cause mortality (in-ICU death or post-discharge death)?} 

\paragraph{Covariates}
The set of covariates $\{L\}$, which are confounders for the decisions and outcomes,  was defined to include variables that plausibly influence both ICU discharge decision-making and the outcome of interest, using clinician expertise and previous literature \citep{knightNurseledDischargeHigh2003}. 
Three baseline covariates were included: age, sex, and admission origin. These variables were treated as fixed throughout the ICU stay and were included to account for baseline patient characteristics and structural differences in care pathways.
In ICU workflow, the patient status is reviewed regularly and discharge readiness is reassessed approximately twice daily. To align the temporal resolution of the data with clinical practice, each ICU admission was discretized into consecutive, non-overlapping windows of 12 hours. Within each window, we have considered 18 time-varying covariates that capture evolving physiological status, laboratory measurements, and respiratory support. Covariates were summarised within each window using either the mean value over the recent past or the most recently observed value, depending on the clinical convention by which the variable is typically interpreted during discharge decision-making. 
The time-varying covariates are summarised in Table \ref{tab:confounders} in appendix \ref{sect:appendix_covariates}, along with their aggregation method, the distribution they follow and more details on covariates processing.

\paragraph{Strategies}
We consider both static discharge strategies apply equally to all patients, and dynamic discharge strategies that recommend discharge conditional on patient history every 12 hours. Under these strategies, ICU discharge is modelled as an absorbing intervention: once discharge occurs, the patient exits the ICU cohort and no further ICU-level interventions are applied.

\subsubsection{Assumption Sanity Checks}
Here we discuss and evaluate the plausibility of the assumptions needed by the parametric g-formula.

\paragraph{Stable Unit Treatment Value (SUTVA)}
This assumption requires (i) well-defined interventions and (ii) no interference between units. 
Regarding (ii), we have assumed that the number of beds in the Beth Israel Deaconess Medical Centre was enough to limit interference (77 according to their website). 
We have included all the stays of patients, yet 18.9\% of the whole discharged population in the MIMIC data are readmitted in the first 90 days, making the admissions not entirely independent. A large part of those readmissions, about 12.24\% of the discharged patients, are readmitted within a day after discharge.
This makes the treatment definition less sharp since some patients actually receive ``extra'' care in a second admission and are not allowed to take the full risk of discharge until the end of follow-up after the first admission. Apart from that, intervention was explicitly defined in section \ref{sect:tte}, and discharge strategies are defined as deterministic rules mapping patient history on the 12-hour grid, constituting a well-defined intervention. 

\paragraph{No unmeasured confounders (NUC)}
Assumptions regarding covariate selection, causal structure, and the specification of discharge strategies were formulated in consultation with ICU clinicians. 
It is formally impossible to test for unmeasured confounders  \cite{schulz_no_2023}. The use of instrumental variables could help \citep{guoUsingInstrumentalVariable2014}, but we did not have access to one and had to rely on clinician expertise. The impact of NUC violation is most important when the unmeasured confounders and the measured ones are uncorrelated \citep{schulz_no_2023}, which, in our case, given the variables included, is arguably unlikely. 

\paragraph{Positivity and Coverage}
There is certainly not enough variability in the data for all the possible strategies to be properly evaluated; indeed, in current care very sick patients are usually kept, and very healthy patients are discharged. 
Focusing on the strategies listed at the end of this section, we applied the positivity checks described in section \ref{sect:assumption}.

\subsubsection{Model Specification Sanity Checks}

To assess the internal validity and plausibility of the estimated effects, we conducted the following diagnostic and sanity checks.

\paragraph{Covariates specification} Distributional assumptions were informed by empirical inspection of variable distributions and selected from the predefined families supported by the \texttt{pygformula} framework, supplemented by exploratory fitting using the \texttt{Fitter} package \citep{cokelaer_fitter}. 
We checked the congruence of the simulation of natural course with the distribution observed in the historical data, by checking Standardized Mean Distance (SMD) variable-wise over time (see figure \ref{fig:SMD_time} in Appendix \ref{sect: appendix_model_sn} and comparisons of empirical distributions in Figures \ref{fig:L1_D1} and \ref{fig:L1_A1}).  

\paragraph{Outcomes specification}
We evaluated whether the g-formula could recover observed outcomes under the natural course strategy. Agreement between estimated and observed post-discharge mortality provides evidence against major model misspecification. Statistically significant deviations may indicate violations of modelling assumptions, unmeasured confounding, or data sparsity at longer ICU stays. To do so, we assumed a Markov-Process where the treatment attribution only depend on current covariates ($f\bigl(\bar{a}^g_{i,t} \mid \bar{\ell}^g_t,\bar{a}^g_t = \bar{1},\bar{y}^g_t = \bar{0} \bigr) = f\bigl(\bar{a}^g_{i,t} \mid \ell^g_t,\bar{a}^g_t = \bar{1},\bar{y}^g_t = \bar{0} \bigr)$). 

\paragraph{Static discharge strategies} 
As a proof of concept, we evaluate simple static discharge strategies under which all patients are discharged after a fixed mount of ICU stay (e.g. after 3 days), irrespective of their clinical status. These strategies are not intended to be clinically meaningful or safe; rather, they serve as a transparent and easily interpretable benchmark. By comparing outcomes under these deterministic time-based rules, which do not consider patient physiology, to those observed under usual care and alternative dynamic strategies, we illustrate that the proposed framework produces sensible estimates. 

\subsubsection{Dynamic strategies evaluation}
We evaluate two dynamic strategies. The first, inspired by \cite{knightNurseledDischargeHigh2003}, requires sustained physiological stability across respiratory, cardiovascular, neurological, and biochemical domains before ICU transfer. Patients are deemed ready for discharge only when predefined thresholds for oxygenation, haemodynamics, neurological status, and key laboratory parameters are met. The complete definition of this strategy is provided in Appendix~\ref{app:knight}. The second is inspired by the recommendation of the guidelines (\citep{nates2016icu}, Table 2 p.1559). Such strategy continues ICU care only when ICU-specific treatments are needed, and otherwise discharges, assuming the patient can be cared for in other wards. 
More precisely, patients are discharged unless predefined physiological red flags indicate severe respiratory, hemodynamic, neurologic or metabolic instability. 
This strategy--referred to as DS1--is detailed in Table \ref{tab:ds1_strategy}.

\begin{table*}[t]
\centering
\small
\setlength{\tabcolsep}{3pt}
\renewcommand{\arraystretch}{1.1}
\begin{tabular}{|p{4.0cm}p{11.5cm}|}
\hline
\textbf{Component} & \textbf{Clinical decision rule (12-hour grid)} \\
\hline



Respiratory failure 
& Invasive controlled ventilation; OR SpO$_2$ $<85\%$; OR O$_2$ flow $>85$; 
OR (ventilation mode unknown AND [SpO$_2$ $<85\%$ OR O$_2$ flow $>85$ OR PaCO$_2$ $>80$ OR respiratory rate $>45$/min]). \\

Hemodynamic instability 
& Mean arterial pressure $<50$ mmHg; OR lactate $>6$ mmol/L; OR heart rate $>160$/min. \\

Neurologic failure 
& Glasgow Coma Scale $<6$. \\

Severe metabolic derangement 
& Arterial bicarbonate $<10$ mmol/L. \\

\hline
\end{tabular}
\caption{Dynamic discharge strategy (DS1) based on physiologic need for organ support: ICU care is continued if any of the red flags above are present, otherwise patients are discharged.}
\label{tab:ds1_strategy}
\end{table*}

\section{Results}
\subsection{Data description}
The study cohort comprised 82,480 ICU admissions from 60,616 unique patients in MIMIC-IV. 
ICU length of stay (administratively truncated at 90 days to align with the follow-up window) had a median of 2.14 days (IQR 1.27-4.11) with a mean of 3.93 days, showing a right-skewed distribution. The distribution of stays were 13\% exceeded 7 days, 4.5\% exceeded 14 days, and 0.8\% exceeded 30 days. 
Overall, 7.38\% of admissions resulted in in-ICU death, while 12.7\% died following ICU discharge within the 90-day follow-up, yielding a composite 90-day all-cause mortality of 20.08\%. Consequently, 92.6\% of admissions were discharged alive from the ICU.


\subsection{Positivity and Coverage}

\begin{figure}[t]
    \centering
    \includegraphics[width=0.45\textwidth]{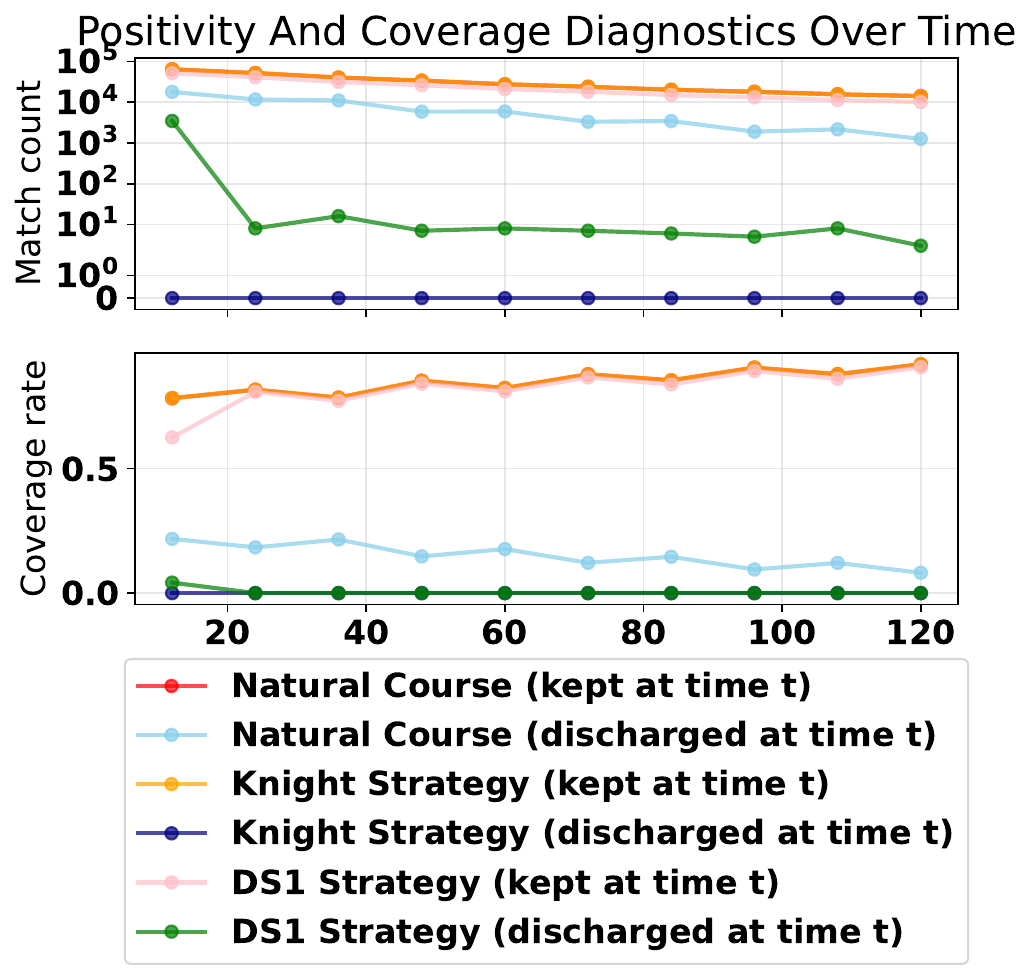}
    \caption{Positivity and coverage diagnostics over time for all evaluated dynamic strategies separated between patients kept and discharged. 
    \textbf{Top:} coverage rate (fraction of history-compatible
    observations at epoch $t$ for which the clinician's action matches the
    target strategy's prescription).
    \textbf{Bottom:} absolute match count (fraction of history-compatible observations at epoch $t$ for which the clinician's action matches the target strategy's prescription and the decision at time $t$).}
    \label{fig:positivity_diagnostics_dynamic}
\end{figure}

The results of such checks are reported in Figure~\ref{fig:positivity_diagnostics_dynamic} and Appendix \ref{sect: appendix_positivity_sn}. In such plots, natural course is the upper bound for match count and coverage rate. 
The static strategy of discharging at day 3 has a match count over 20,000 and coverage around 0.8 until the time where it discharges patients. At discharge time, metrics deteriorates (coverage around 0.2, match count of 3,321 for discharge at 72h) (Figures \ref{fig:positivity_diagnostics} and \ref{fig:positivity_pca_static}).

Both dynamic strategies seem to have good results for these two metrics: over 10,000 match count and 0.6 coverage (Figure~\ref{fig:positivity_diagnostics}). Yet, when splitting depending on the decision at time $t$, we observe that no patients are discharged under Knight strategy (match count of 0) and only  few with the DS1 strategy (match count of 3,476 the first 12 hours then less than 10) (Figure~\ref{fig:positivity_diagnostics_dynamic}). This was missed by the first metrics due to the fact that match count and coverage were drowned by the kept patients, for which Knight and DS1 strategy are close to natural course (match count over $10^3$ and coverage around 0.8). This amount of data hides the support issue. Visualizations after application of PCA further support this conclusion (Figure \ref{fig:positivity_pca_dynamic}).

\subsection{Checks on the simulation}

\begin{table*}[t]
\centering
\small
\setlength{\tabcolsep}{1pt}
\renewcommand{\arraystretch}{1.0}
\begin{tabular}{|l||ccc|cc||cc|ccc|}
\hline
& \multicolumn{3}{|c|}{\textbf{All-cause mortality (\%)}}
&\multicolumn{2}{|c||}{\textbf{Difference to}} 
&\multicolumn{2}{c|}{\textbf{Mortality split (\%)}}
& \multicolumn{3}{|c|}{\textbf{Length of ICU stay (days)}}\\  
\textbf{Intervention}
& Observed  
&\multicolumn{2}{c|}{G-formula estimate} 
&\multicolumn{2}{|c||}{\textbf{natural course (\%)}} 
& In ICU & After discharge 
&\multicolumn{2}{c}{Median (IQR)} & Mean\\ 
\hline

Natural course 
& 20.08 
& 19.1
& (18.4, 21.5) 
& -- 
& -- 
& 6.2 & 12.9 & 2.00 &(1.00--4.00) & 3.44 \\


Discharge 3rd day 
& -- 
& 17.4 
& (16.7, 20.1) 
& -1.8 
& (-3.3, 0.5) 
& 5.2 & 12.2 &3&(3--3)&3\\

Strategy DS1
& -- 
& 18.0 
& (15.8, 18.9) 
& -1.2 
& (-4.4, -0.6) 
& 3.6 & 14.4 & 0.5 & (0.5--1.0) & 1.76 \\

Strategy \citep{knightNurseledDischargeHigh2003} 
& -- 
& 86.0
& (83.4, 89.4)
& +66.9
& (63.5, 69.5)
& 85.5 & 0.5 & 35.0 &(15.5--62.5) & 40.56\\

\hline
\end{tabular}
\caption{Estimated 90-day all-cause mortality (in-ICU or post-discharge death) under dynamic and static discharge strategies using the parametric g-formula (NICE estimator). Mean and 95\% confidence intervals are shown. Follow-up corresponds to 180 half-day intervals (90 days).}
\label{tab:gformula_results}
\end{table*}


\paragraph{Sanity check on covariate distribution}
Across both clinical endpoints discharge and in-ICU death, the natural course simulation closely reproduces the empiricial covariate distributions observed in MIMIC-IV, as shown in Figures \ref{fig:L1_A1} and \ref{fig:L1_D1}. At discharge, physiologic variables reflecting clinical stability (heart rate, mean arterial pressure, respiratory rate, temperature, hemoglobin, bicarbonate, and arterial blood gases) exhibit almost complete overlap of distributions. Oxygenation measures (SpO2, PaO2, O2 flow), renal markers (creatinine, ureum, urine output), and lactate similarly show highly right-skewed distributions. At in-ICU death, where the physiology deranges severely (e.g., elevated lactate and creatinine, broader respiratory distributions, lower GCS), the simulated natural course preserves both the heavier tails and distributional shifts seen in the observed data. The categorical distribution of ventilation model is also well aligned at both end points. The consistency of these high-dimensional distributional matches across both states providing empirical evidence that the natural course model has successfully replicated the implicit strategy in the MIMIC-IV data.

\paragraph{Natural Course Validation} Under the natural course (observed discharge strategy in MIMIC-IV), the observed or true 90-day mortality was 20.08\%. The parametric g-formula estimate under the natural course was 19.1\% with a 95\% CI 18.4\%-21.5\%. The observed value lying within the confidence interval indicates good internal calibration of the models for time-varying covariates, discharge decisions, in-ICU death, and post-discharge mortality. This supports the adequacy of model specification. Under the observed natural course, 7.38\% of patients died in the ICU and 12.9\% died after discharge, corresponding to approximately 38.6\% and 61.4\% of all deaths, respectively.

\paragraph{Static strategy (Discharge on 3rd day in the ICU)} We evaluated a static intervention that discharges all patients on ICU day 3 (12-hour grid). Under this strategy, the estimated 90-day mortality was 17.4\% (95\% CI: 16.7\%-20.1\%). Compared with natural course, this corresponds to a risk reduction of 1.8 percentage points (95\% CI: -3.3 to 0.5) as reported in Table \ref{tab:gformula_results}. The reduced in-icu mortality of 5.2\% is in accordance to an early discharge. The post-discharge mortality only marginally decreases by 0.7 percentage points. Although the point estimate suggests lower mortality under day-3 discharge, the confidence interval for the mean difference includes the null, indicating statistical uncertainty regarding the presence and direction of effect. 


\paragraph{Dynamic strategy Knight}
Under the Knight discharge strategy, the estimated 90-day mortality was 86.0\% (95\% CI: 83.4\%–89.4\%), corresponding to an absolute risk increase of 66.9 percentage points relative to the natural course (95\% CI: 63.5 to 69.5). The confidence interval excludes the null, indicating a statistically significant and substantial increase in mortality under this strategy. 

Decomposition of mortality shows that 85.5\% of patients died in the ICU and only 0.5\% after discharge, indicating that nearly all deaths occurred inside the ICU. The strategy is highly conservative, as discharge requires all physiological criteria to be simultaneously satisfied, resulting in markedly prolonged ICU stays (median 35.0 days [IQR 15.5–62.5]; mean 40.56 days). This extended exposure to ICU-level severity is reflected in the extreme concentration of deaths within the ICU and explains the increase in overall mortality under this strategy.

\paragraph{Dynamic strategy DS1}
The dynamic strategy DS1 yielded an estimated 90-day mortality of 18.0\% (95\% CI: 15.8\%-18.9\%). Relative to the natural course, this corresponds to an absolute risk reduction of 1.2 percentage points (95\% CI: -4.4 to -0.6). The point estimate suggests lower overall mortality under the DS1 discharge strategy that balances physiological stability with conditions manageable outside the ICU. The confidence interval for the mean difference excludes zero, indicating a statistically significant reduction in mortality compared with the natural course. 

Mortality decomposition shows that in-ICU deaths decreased to 3.6\% (from 6.2\% under natural course), whereas post-discharge mortality increased slightly to 14.4\% (from 12.9\%). This pattern indicates partial redistribution of deaths from the ICU to the post-discharge period. The strategy was also associated with substantially shorter ICU stays (median 0.5 days [IQR 0.5–1.0]; mean 1.76 days), as expected from a more ``aggressive'' discharge strategy.


\section{Discussion}
In this paper we proposed a framework to evaluate strategies to discharge patients from the ICU, pooling together clinical expertise with techniques from Causal Inference and Reinforcement Learning. Our implementation modified the existing pygformula package to allow handling of scenarios where stopping the intervention prevents us from observing covariate evolution. We then suggested a way to appraise the estimate of potential outcome by running positivity and coverage checks. Finally, we showcased this methodology by applying it to real-world ICU data and tested a few clinically-motivated dynamic strategies. Our results chart a way forward in solving this problem but also underscore the complexity of off-policy evaluation in the ICU environment.

\paragraph{Strengths.} The code used in this paper is fully open source, as is the dataset, hence our results are fully replicable and can function as a useful starting point for further studies. Both our workflow and our results are vetted by clinicians, ensuring the plausibility of modelling decisions, e.g. discharge strategies to consider.
Furthermore, while our use case is clinical, the math and the code are easily generalizable to any other scenario with a similar structure.

\paragraph{Limitations.} Our results are also subject to some limitations. On the side of the estimation, the identification of effects via the g-formula requires the standard assumptions of SUTVA, positivity and no unmeasured confounders. The NICE estimator we employed requires causal mechanisms to be stable across time steps; we also assumed all models used within NICE to be generalized linear. Finally, our landmarking approach for estimating post-stopping outcomes assumes that we can obtain unbiased estimates of said outcome with only covariates at the time of stopping and information concerning time itself. 

The last assumption in particular is what needs to be added when employing our approach to handle missing covariates after the stopping decision is taken. In the case of ICU discharge, this assumption may not hold if for instance there are unmeasured confounders influencing the quality of care patients receive post discharge. When inspecting the outcome models, we find the post-discharge estimates a tad too flat across strategies and the in-ICU estimates too extreme for long stays (as seen for the Knight strategy), which suggest that more work on the modelling might be needed.

We also made some simplifications on the data processing side. While our representations of ICU patient are quite rich, they still did not contain information e.g. on diagnoses or medications--although they did contain information about settings of mechanical ventilation. Patients were also not `allowed' to be readmitted in the ICU once discharged. Moreover, our discrete-time view on the data generation process and related simulation assumes a sequential flow (first check if patient is alive at this time step, then generate covariates, then make stopping decision) which may over-simplify reality.

In light of these assumptions, despite having found a strategy with statistically significant improvement over natural course, we stress that the results require more thorough clinical investigation before being used to influence clinical care. In particular, clinicians need to be involved when assessing the desired trade-off between mortality outcome and length of stay in the ICU.
This disclaimer notwithstanding, there are some checks reassuring us that there are no gross violations in the capturing of the causal mechanisms. One of them is the verification that the simulation of natural course is able to ``reconstruct'' the observed data. 

Concerning the larger goal of finding an optimal stopping strategy for the scenarios we are interested in, our work provides a first building block for algorithmic search of optimal strategy, insofar as this process can be conceived as an ``oracle'' returning the potential outcome given a strategy $g$ (keeping fixed the observational dataset). Future work in this direction will investigate how such search can be performed in presence of obstacles such as non-differentiable threshold-based strategies and discrete variables in the simulations.

\bibliography{uai2026-template}

\newpage

\onecolumn

\title{A Causal Framework for Evaluating ICU Discharge Strategies\\(Supplementary Material)}
\maketitle

\appendix
\section{Method appendix}

\subsection{Mathematical formulation of causal assumptions}\label{sect:appendix_assumptions}
The assumptions below correspond to the one of \cite{deliuDynamicTreatmentRegimes2022} for DTR:
\begin{itemize}
    \item \textbf{Stable Unit Treatment Value (SUTVA):} $Y^{obs}_{i,t} = \sum_{\bar{a}_{i,t}} Y_{i,t}(\bar{a}_{i,t}) \mathbb{I}\{\bar{a}^{obs}_{i,t} = \bar{a}_{i,t}\}$ implying no interference between units and a well-defined version of each treatment,
    \item \textbf{No Unmeasured Confounders (NUC):} $Y_{t}(\bar{A}_{t}) \perp \bar{A}^{obs}_{t}| X, \bar{H}_t$ implying that we have measure all the necessary variables to recover the unknown strategy of the data, and to guaranty the idependence between treatment allocation and the potential outcomes,
    \item \textbf{General positivity:} $0 < f\bigl(\bar{a}_{t} \mid \bar{\ell}_t,\bar{a}_t = \bar{1},\bar{y}^g_t = \bar{0} \bigr)$, $\forall t, \bar{a}_{t} \in \bar{A}_{t}, \bar{\ell}_{t} \in \bar{L}_{t}$ meaning that there are enougth variability in the unknown strategy so that any possible history has a positive probability of receiving either treatment,
    \item \textbf{Positivity for the strategy $g$:} $0 < f\bigl(a^g_{t}\mid \bar{\ell}^g_t,\bar{a}^g_t = \bar{1},\bar{y}^g_t = \bar{0} \bigr)$, $\forall t, \bar{\ell}^g_{t} \in \bar{L}^g_{t}$ meaning that there are enougth variability in the data so that for any possible history under strategy $g$ at time $t$ there are some patient with that history that follow the decision at time $t$.
\end{itemize}

\subsection{History Modeling }
\label{sect:history}
As we assume a stable Markov process, we need to summarise in a constant way the time-growing history for each density modelled, making the following assumptions:
\begin{itemize}
    \item for the covariates density:
    \begin{align*}
        f\bigl(L^g \mid \bar{\ell}^g_t,\bar{a}^g_t = \bar{1},\bar{y}^g_t = \bar{0} \bigr) = f\bigl(L^g \mid \ell^g_t,t,\bar{y}^g_t = \bar{0} \bigr)
    \end{align*}
    \item for the in-ICU outcome density:
    \begin{align*}
        f\bigl(Y^g \mid \bar{\ell}^g_t, \bar{a}_t^g = \bar{1}, \bar{y}_t^g= \bar{0}\bigr) = f\bigl(Y^g \mid \ell^g_{t-1},\ell^g_{t-2},t, \bar{y}_t^g= \bar{0}\bigr)
    \end{align*}
    if $t = 1$ and thus $\ell^g_{t-2}$ is not defined, we put NaNs.
    \item for the discharged outcome density:
    \begin{align*}
        f\bigl(Y^g \mid \bar{\ell}^g_{\tau_i^g}, \bar{a}_t, \bar{y}_t^g= \bar{0} \bigr) = f\bigl(Y^g \mid \frac{1}{\tau_i^g + 1}\sum_{j = 0}^{\tau_i^g}\ell^g_{j},\tau_i^g, t - \tau_i^g, \bar{y}^g= \bar{0} \bigr)
    \end{align*}
    \item Natural course treatment probability:
    \begin{align*}
        f\bigl(\bar{a}^g_{i,t} \mid \bar{\ell}^g_t,\bar{a}^g_t = \bar{1},\bar{y}^g_t = \bar{0} \bigr) = f\bigl(\bar{a}^g_{i,t} \mid \ell^g_t,\bar{a}^g_t = \bar{1},\bar{y}^g_t = \bar{0} \bigr)
    \end{align*}
\end{itemize}

\subsection{Covariates description}\label{sect:appendix_covariates}

\subsubsection{Baseline covariates}
Baseline covariates were defined using measurements recorded within the first 5 hours of ICU admission, reflecting the patient’s initial physiological state prior to discharge decision-making. These included demographics (age, sex), admission origin. Other baseline covariates such as height, weight and ICU unit type were omitted due to significant missingness.

\subsubsection{Time-varying covariates}
To align the temporal resolution of the data with clinical practice, each ICU admission was discretised into consecutive, non-overlapping windows of 12 hours. Within each 12-hour window, time-varying confounders capture evolving physiological status, laboratory measurements, and respiratory support. The time-varying confounder set comprised arterial bicarbonate, activated partial thromboplastin time, mean temperature, haemoglobin, mean heart rate, mean arterial blood pressure, creatinine, urea, mean urine output, lactate, Glasgow Coma Scale total score, mean arterial pCO$_2$, mean respiratory rate, mean oxygen saturation, oxygen flow rate, mean arterial pO$_2$, ventilator mode, and time since last ventilator mode recorded. Ventilation mode, originally recorded using 42 granular categories in the source data, was mapped to four clinically meaningful groups: \textit{unknown}, \textit{invasive\_controlled}, \textit{invasive\_assisted}, and \textit{cancelled} based on expert input from ICU clinicians. This consolidation reflects how ventilatory support is operationally assessed in practice and reduces sparsity while preserving clinically relevant distinctions for modelling discharge decisions and outcome.

The method used to summarise the variables over 12-hour windows and the distribution used for each variable in the simulations are summarised in Table \ref{tab:confounders}. 

\begin{table}[h!]
\centering
\small
\begin{tabular}{|l|l|l|}
\hline
\textbf{Variable name} & \textbf{Summary over last 12h} & \textbf{Distribution }\\

\hline
Ventilation mode & last status & categorical \\
Hours since last recorded vent mode & last value & zero-inflated normal\\
Arterial pco2 & mean & truncated normal\\
Arterial po2\ & mean & bounded normal\\
Oxygen flow & last value & bounded normal\\
O2 saturation & mean & bounded normal\\
Respiratory rate & mean & bounded normal\\
Glasgow coma scale & last value & bounded normal\\
Lactate & last value & truncated normal\\
Fluid out urine & mean & zero-inflated normal\\
Ureum & last value & truncated normal\\
Creatinine & last value & truncated normal\\
Mean Arterial blood pressure & mean & bounded normal\\
Heart rate & mean & bounded normal\\
Hemoglobin & last value & bounded normal\\
Mean Temperature & mean & normal\\
Activated partial thromboplastin time & last value & truncated normal\\
Arterial bicarbonate & last value & bounded normal\\
\hline
\end{tabular}
\caption{Clinical variables with their 12-hour aggregation type and distribution}
\label{tab:confounders}
\end{table}

\subsection{Description of strategies}
\label{app:knight}
The Knight discharge strategy \citep{knightNurseledDischargeHigh2003} operationalizes 15 physiological and laboratory stability criteria across respiratory, cardiovascular, neurological, and biochemical categories, each required to be within predefined thresholds. ICU discharge is recommended only when all criteria are satisfied, indicating sustained clinical stability.

\begin{table}[t]
\centering
\small
\setlength{\tabcolsep}{6pt}
\renewcommand{\arraystretch}{1.15}
\begin{tabular}{lll}
\hline
\textbf{Test name} & \textbf{Variable} & \textbf{Test condition} \\
\hline
Respiratory: airway & airway & airway patent \\
Respiratory: FiO$_2$ & fio2 & fio2 $\leq$ 0.6 \\
Respiratory: blood oxygen & spo2 & spo2 $\geq$ 95 (\%) \\
Respiratory: bicarbonate & hco3 & hco3 $\geq$ 19 (mmol/L) \\
Respiratory: rate & resp (rate) & 10 $\leq$ resp $\leq$ 30 (bpm) \\
Cardiovascular: blood pressure & bp (systolic) & bp $\geq$ 100 (mm Hg) \\
Cardiovascular: heart rate & hr & 60 $<$ hr $\leq$ 100 (bpm) \\
Pain & pain & 0 $\leq$ pain $\leq$ 1 \\
Central nervous system & gcs & gcs $\geq$ 14 \\
Temperature & temp & 36 $\leq$ temp $\leq$ 37.5 ($^\circ$C) \\
Blood: haemoglobin & haemoglobin & haemoglobin $\geq$ 90 (g/L) \\
Blood: potassium & k & 3.5 $\leq$ k $\leq$ 6.0 (mmol/L) \\
Blood: sodium & na & 130 $\leq$ na $\leq$ 150 (mmol/L) \\
Blood: creatinine & creatinine & 59 $\leq$ creatinine $\leq$ 104 ($\mu$mol/L) \\
Blood: urea & bun & 2.5 $\leq$ bun $\leq$ 7.8 (mmol/L) \\
\hline
\end{tabular}
\caption{Codified discharge criteria applied to electronic health record data. All criteria must be satisfied prior to ICU discharge.}
\label{tab:knight_discharge_criteria}
\end{table}



\section{Positivity and coverage sanity check}
\label{sect: appendix_positivity_sn}

\subsection{Additional Method}


\subsubsection{Importance-sampling and Effective Sample Size Ratio}

We define the set of history-compatible trajectories at
epoch $t$, $\mathcal{D}^g_t$, from which can be extracted a set of trajectories for which the decision at time $t$ coincide with strategy $g$, $\mathcal{M}^g_t$. We denote $\bar{h}^g_{t-1} = (\bar{\ell}^g_t,\bar{a}^g_t = \bar{1},\bar{y}^g_t = \bar{0})$ the history at time $t$.

\paragraph{Importance-sampling} (IS) weights were design to identify when overlap assumption is not fullfilled:
\begin{align*}
w(g,a, \bar{h}_{t-1}) = \frac{f\bigl(a^g_{t} = a \mid \bar{h}^g_{t-1} = \bar{h}_{t-1} \bigr)}{f\bigl(a^{NC}_{t} = a \mid \bar{h}^{NC}_{t-1} = \bar{h}_{t-1} \bigr)}
\end{align*}
Indeed, they are undefined when the probability of taking a certain action in natural course is null. 

\paragraph{Deterministic target strategies.}
The strategies evaluated here—both the static and the dynamic ones—are \emph{deterministic}: at each
decision epoch $t$ and for each patient trajectory, the strategy prescribes
exactly one action (keep or discharge) with
probability one.
Positivity therefore reduces to a binary feasibility check: every state
in which $g$ prescribes discharge must have been observed with at least
one clinician-initiated discharge, and vice versa for keep:
\begin{align*}
w(g,a, \bar{h}_{t-1}) = \frac{\mathbb{I}\{a^g_{t} = a\}}{f\bigl(a^{NC}_{t} = a \mid \bar{h}^{NC}_{t-1} = \bar{h}_{t-1} \bigr)}
\end{align*}
where $\mathbb{I}$ is the indicator function. A \emph{positivity violation} at $(\bar{h}_{t-1}, a)$ means that the strategy dictates
action $a$ in a region of state space where clinicians \emph{never} took that
action, making any estimate for trajectories passing through that region unreliable.

\paragraph{Coefficient of variation} The coefficient of variation, further noted CV correspond to the variance of the weights normalised by the mean, assuming this mean is different from zero:
\begin{align*}
\text{CV}_{i \in \mathcal{M}_t}(\hat{w}(g,a^{NC}_{i,t}, \bar{h}^{NC}_{i,t-1})) = \frac{V_{i \in \mathcal{M}_t}(\hat{w}(g,a^{NC}_{i,t}, \bar{h}^{NC}_{i,t-1}))}{E_{i \in \mathcal{M}_t}(\hat{w}(g,a^{NC}_{i,t}, \bar{h}^{NC}_{i,t-1}))^2}
\end{align*}
When used with the importance sampling weights, a coefficient of variation near zero means that all the action given histories have the same importance sampling, whereas when the CV is high, some might be overrepresented and others underrepresented. Yet, this measure gives a sense of the variability of the weights but cannot be interpreted in terms of sample size power.

\paragraph{Effective sample size ratio} A usefull metric to give a sense of the sample size reduction linked to the emballenced sampling in the effective sample size :
\begin{align*}
    \mathrm{ESS}_t &= \frac{\left(\sum_{i \in \mathcal{M}_t} \hat{w}(g,a^{NC}_{i,t}, \bar{h}^{NC}_{i,t-1})
    \right)^2}{\sum_{i \in \mathcal{M}_t} \hat{w}(g,a^{NC}_{i,t}, \bar{h}^{NC}_{i, t-1})^2} 
    = \frac{|\mathcal{M}_t|}{1+\text{CV}^2_{i \in \mathcal{M}_t}(\hat{w}(g,a^{NC}_{i,t}, \bar{h}^{NC}_{i,t-1}))}
    \label{eq:ess}
\end{align*}
It corresponds to the number of balanced sampled observations that would be needed, so that the variance of the mean weighted by the importance sampling weights is equal to the variance of the unweighted mean.
\begin{align*}
    V(\frac{1}{|\mathcal{M}_t|}\sum_{i \in \mathcal{M}_t} Y_i) = V(\sum_{i \in \mathcal{M}_t} \hat{w}(g,a^{NC}_{i,t}, \bar{h}^{NC}_{i,t-1})Y_i)
\end{align*}
A value close to $|\mathcal{M}_t|$ means that the action given the history was homogenously sampled, whereas a value close to 0 means that only one type of action given history was over sampled compare to the rest of them. This metric thus enable to identify an underbalanced sampling, which could lead to near positivity violation. The ratio enables a sense of what proportion of the data contains actual non-redundant information.

\subsubsection{PCA visualization}

Finally, to ground our intuition, we also visualize the areas of the feature space in which the two strategies disagree. After reducing the dimensionality of the dataset to 3 via PCA, we colour-code the areas in which the decision of the strategy $g$ agreed with the natural course (orange for discharge, blue for keep) while we mark in grey points where the is complete disagreement, i.e. positivity violations.

\subsection{Additional Results}

\begin{figure}[h!]
    \centering
    \includegraphics[width=\textwidth]{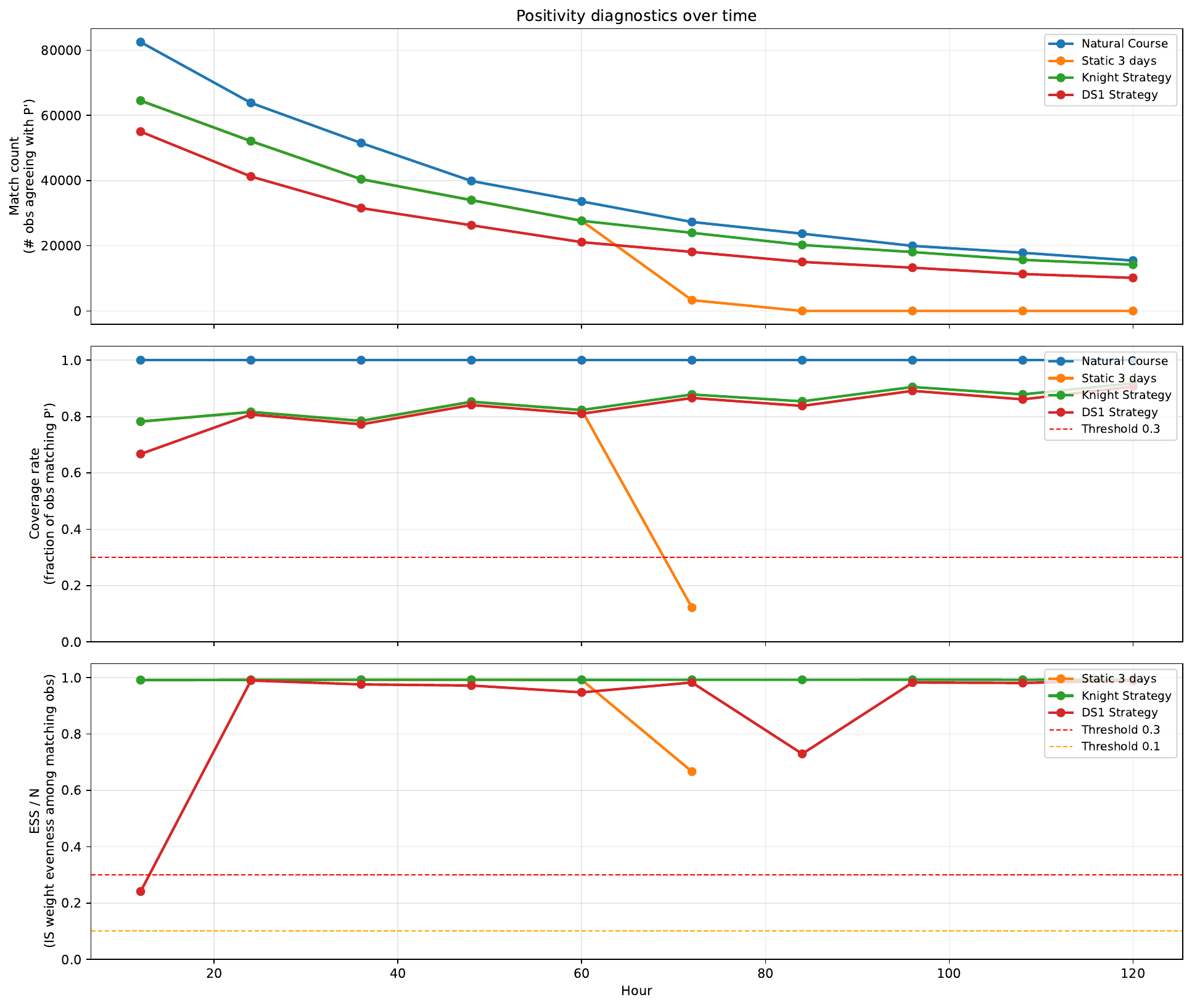}
    \caption{Positivity diagnostics over time for all evaluated strategies.
    \textbf{Top:} coverage rate $\rho_t$ (fraction of history-compatible
    observations at epoch $t$ for which the clinician's action matches the
    target strategy's prescription).
    \textbf{Middle:} effective sample size ratio $\mathrm{ESS}_t / N_t$
    among matching observations (IS weight evenness; 1 = uniform, $1/N_t$
    = fully concentrated).
    \textbf{Bottom:} absolute match count $N_t$.
    The red dashed line in the top and middle panels marks the 0.3 warning
    threshold.}
    \label{fig:positivity_diagnostics}
\end{figure}

\begin{figure}[h!]
    \centering
    \foreach \strategy in {Static_3_days}{%
        \includegraphics[width=\textwidth]{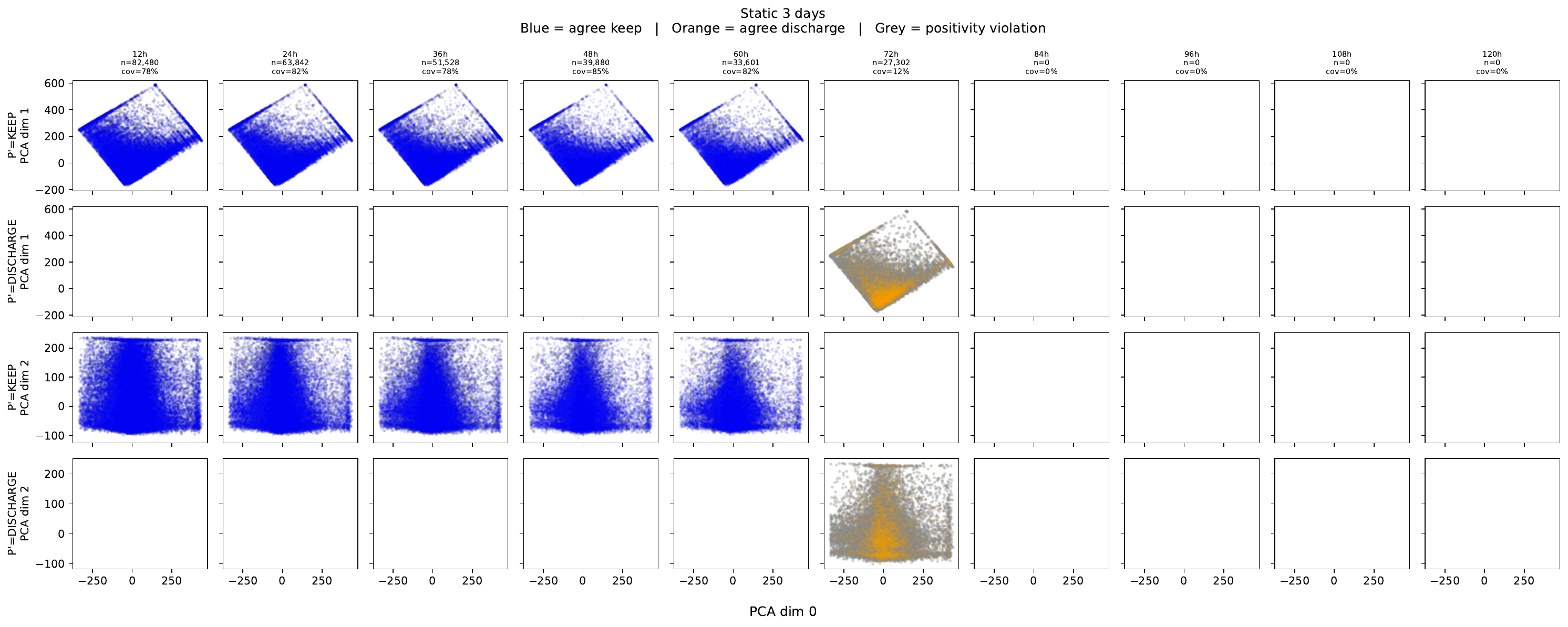}%
        \par\smallskip
    }
    \caption{PCA positivity scatter plots for the three static threshold
    strategy at 3 days, across decision
    epochs $t \in \{12, 24, \ldots, 120\}$\,h (columns).
    Each panel projects the first two PCA components of the covariate space.
    \textbf{Rows 1 \& 3}: blue dots indicate patients
    for whom the strategy prescribes keep \emph{and} the clinician kept;
    grey dots indicate patients for whom the strategy prescribes keep but the
    clinician discharged (positivity violation).
    \textbf{Rows 2 \& 4}: orange dots indicate
    agreement on discharge; grey dots indicate patients the strategy would
    discharge but clinicians retained (positivity violation).
    Panel titles report cohort size $n$ and coverage rate $\rho_t$.}
    \label{fig:positivity_pca_static}
\end{figure}

\begin{figure}[h!]
    \centering

    \begin{subfigure}{\textwidth}
        \centering
        \includegraphics[width=\textwidth]{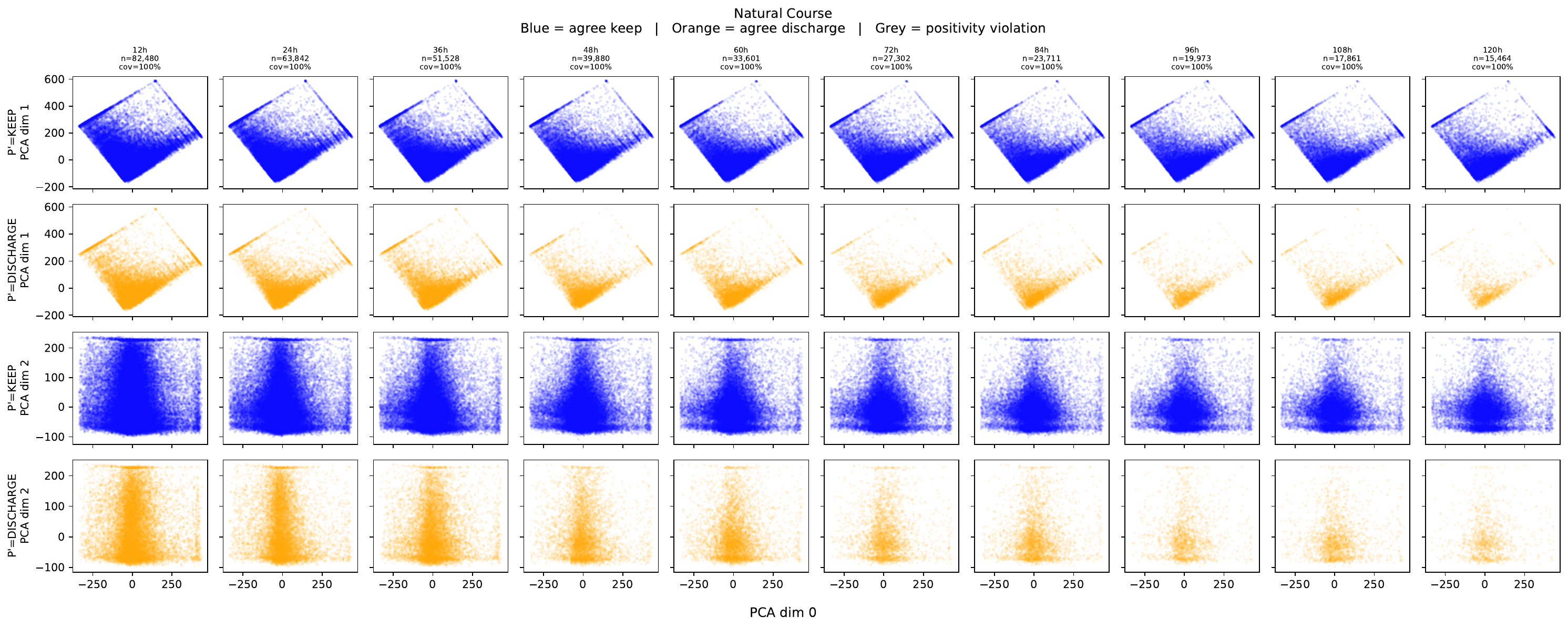}
    \end{subfigure}
    \par\smallskip

    \begin{subfigure}{\textwidth}
        \centering
        \includegraphics[width=\textwidth]{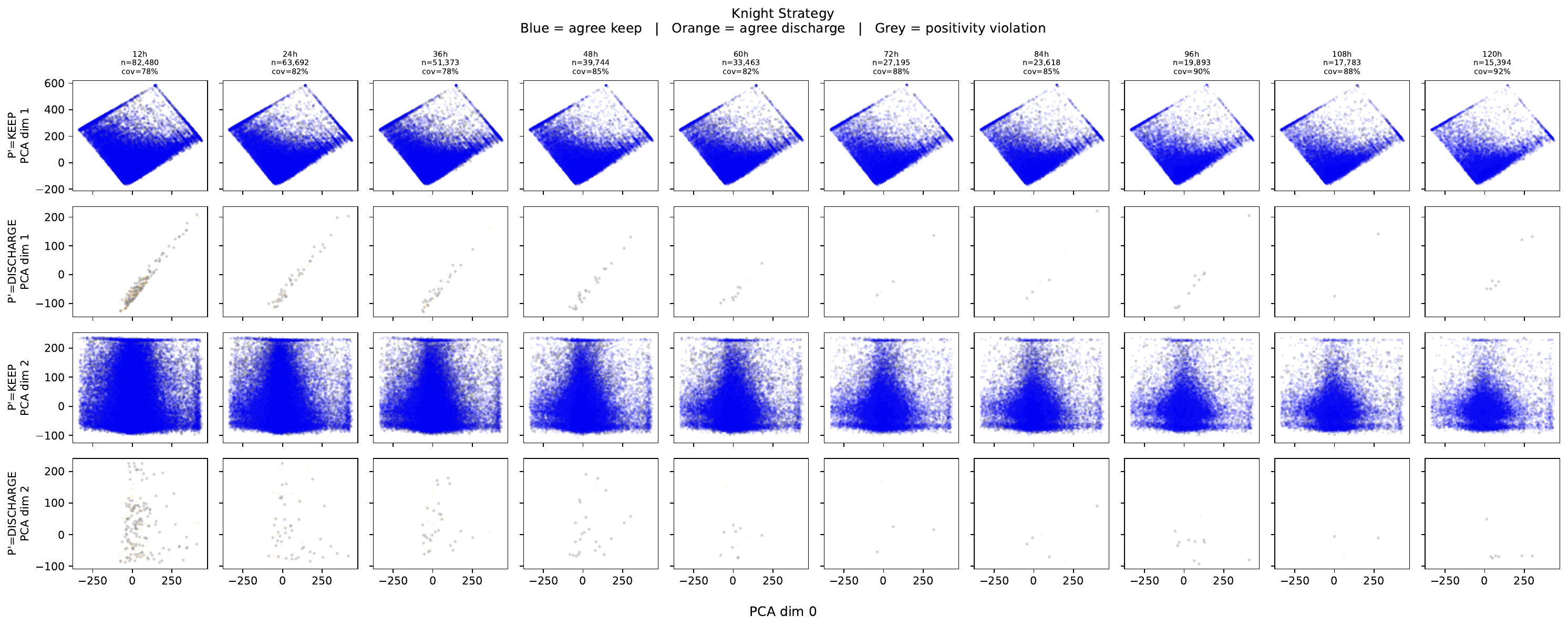}
    \end{subfigure}
    \par\smallskip

    \begin{subfigure}{\textwidth}
        \centering
        \includegraphics[width=\textwidth]{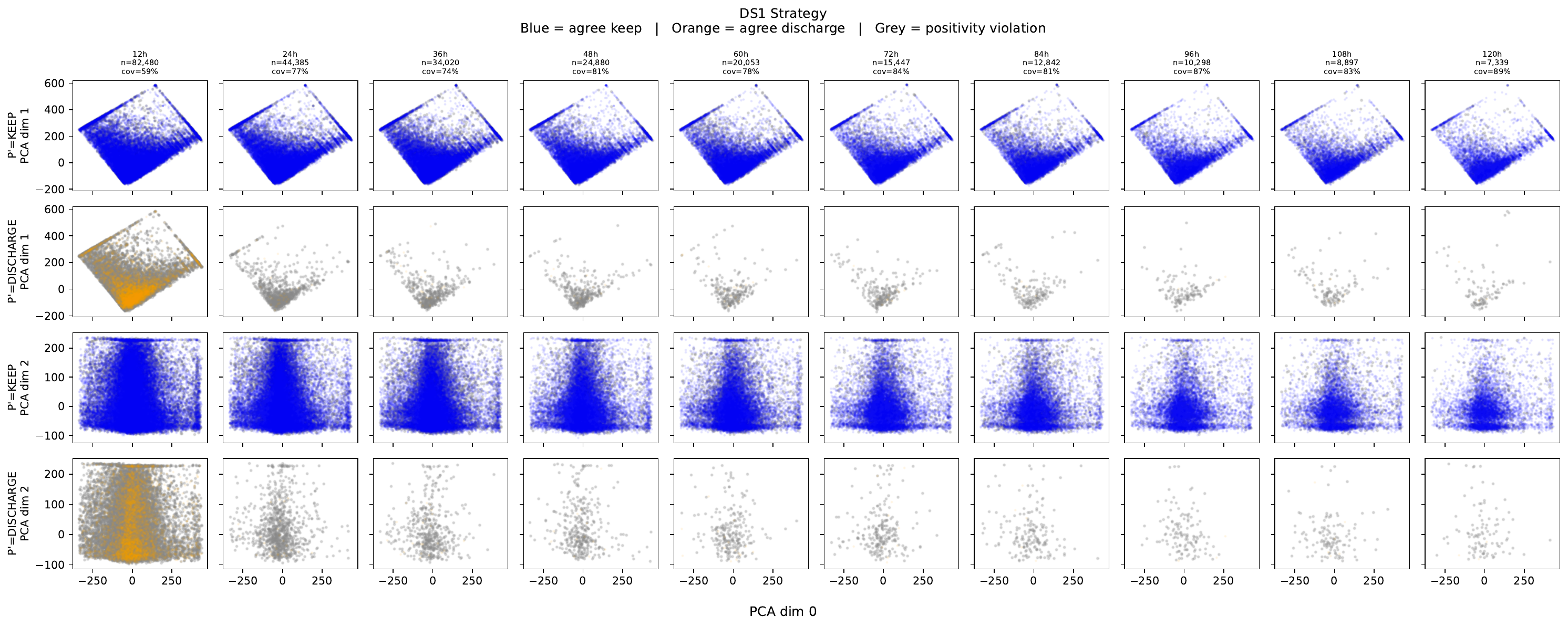}
    \end{subfigure}

    \caption{PCA positivity scatter plots for the dynamic strategies
    (Natural course, Knight strategy, and DS1 strategy; top to bottom),
    across decision epochs $t \in \{12, 24, \ldots, 120\}$\,h (columns).
    Layout and colour coding are identical to Figure~\ref{fig:positivity_pca_static}.}

    \label{fig:positivity_pca_dynamic}
\end{figure}

\clearpage
\section{Model Specification Sanity Checks}\label{sect: appendix_model_sn}
\subsection{Natural Course Calibration via Standardized Mean Differences}
\begin{figure*}[t]
    \centering
    \includegraphics[width=1.15\textwidth]{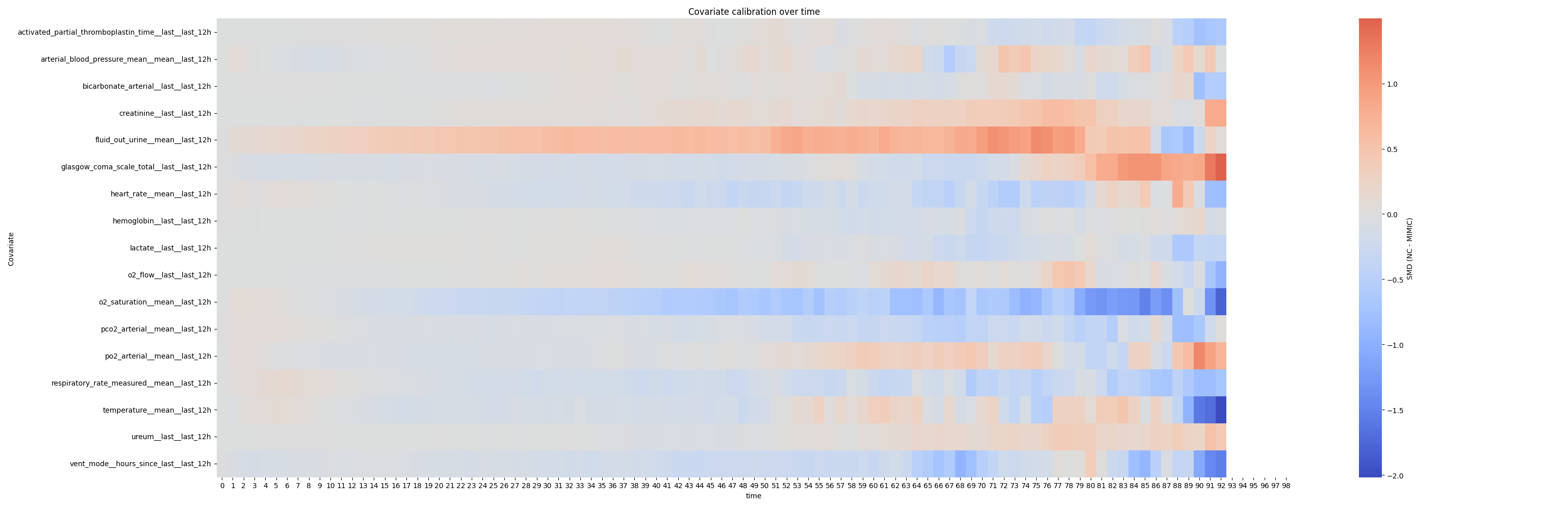}
    \caption{Time-indexed standardized mean differences comparing observed and simulated covariate evolution under the natural course.}
    \label{fig:SMD_time}
\end{figure*}

As a primary internal validation step, we evaluated whether simulated covariate trajectories under the natural course reproduced the joint distribution observed in the historical data. Figure \ref{fig:SMD_time} presents time-indexed standardized mean differences (SMDs) for each covariate, comparing observed and simulated values over the follow-up. Small SMDs over time indicate adequate calibration of the longitudinal covariate models. Larger SMDs point to potential model misspecifications.

\subsection{Empirical Distribution Diagnostics}
\begin{figure*}[t]
    \centering
    \includegraphics[width=\textwidth]{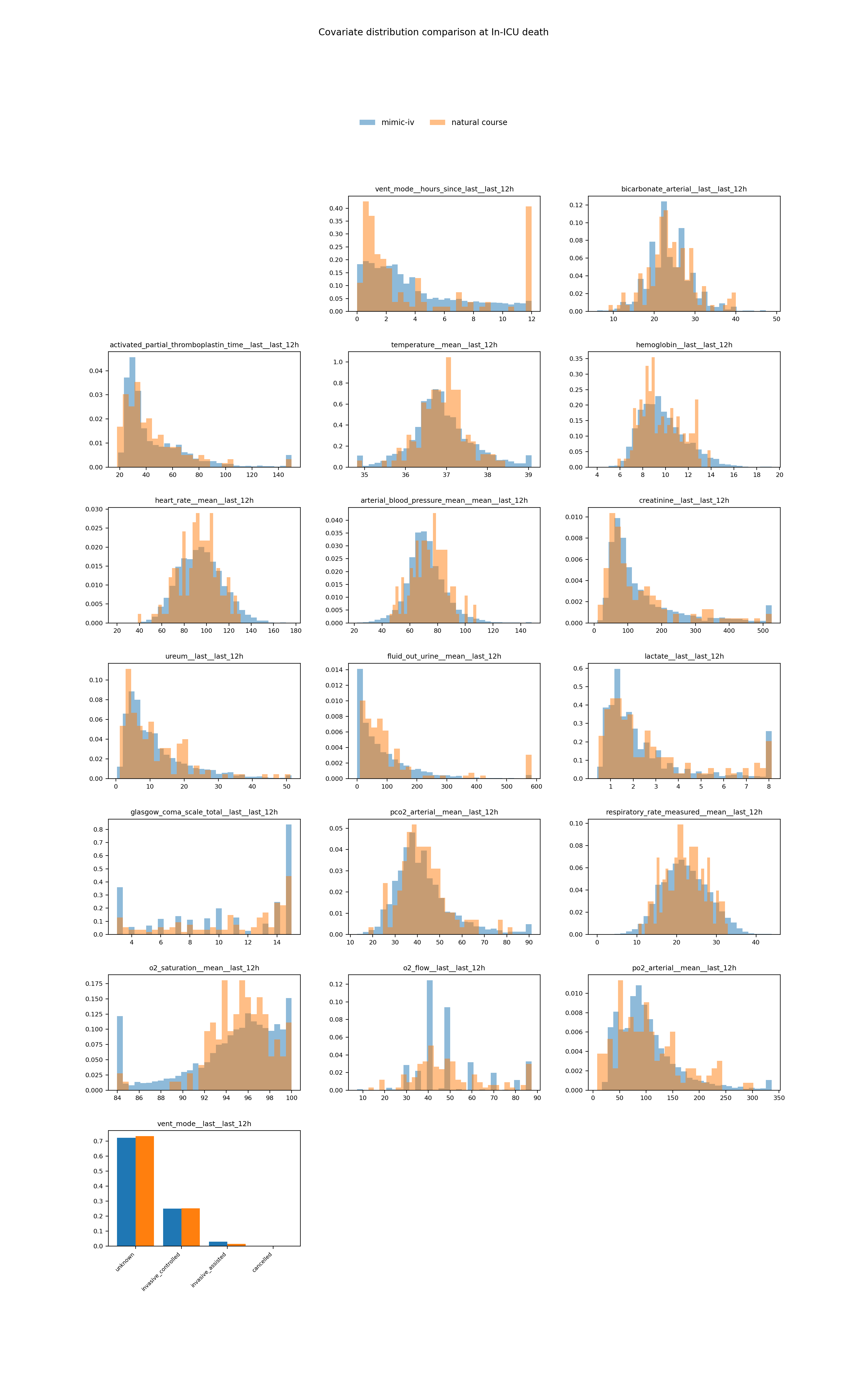}
    \caption{Covariate distribution comparison under the natural course and MIMIC-IV data for In-ICU death.
    The close overlap across physiologic variables demonstrates that the natural course simulation reproduces the empirical covariate distributions at In-ICU death.}
    \label{fig:L1_D1}
\end{figure*}

\begin{figure*}[t]
    \centering
    \includegraphics[width=\textwidth]{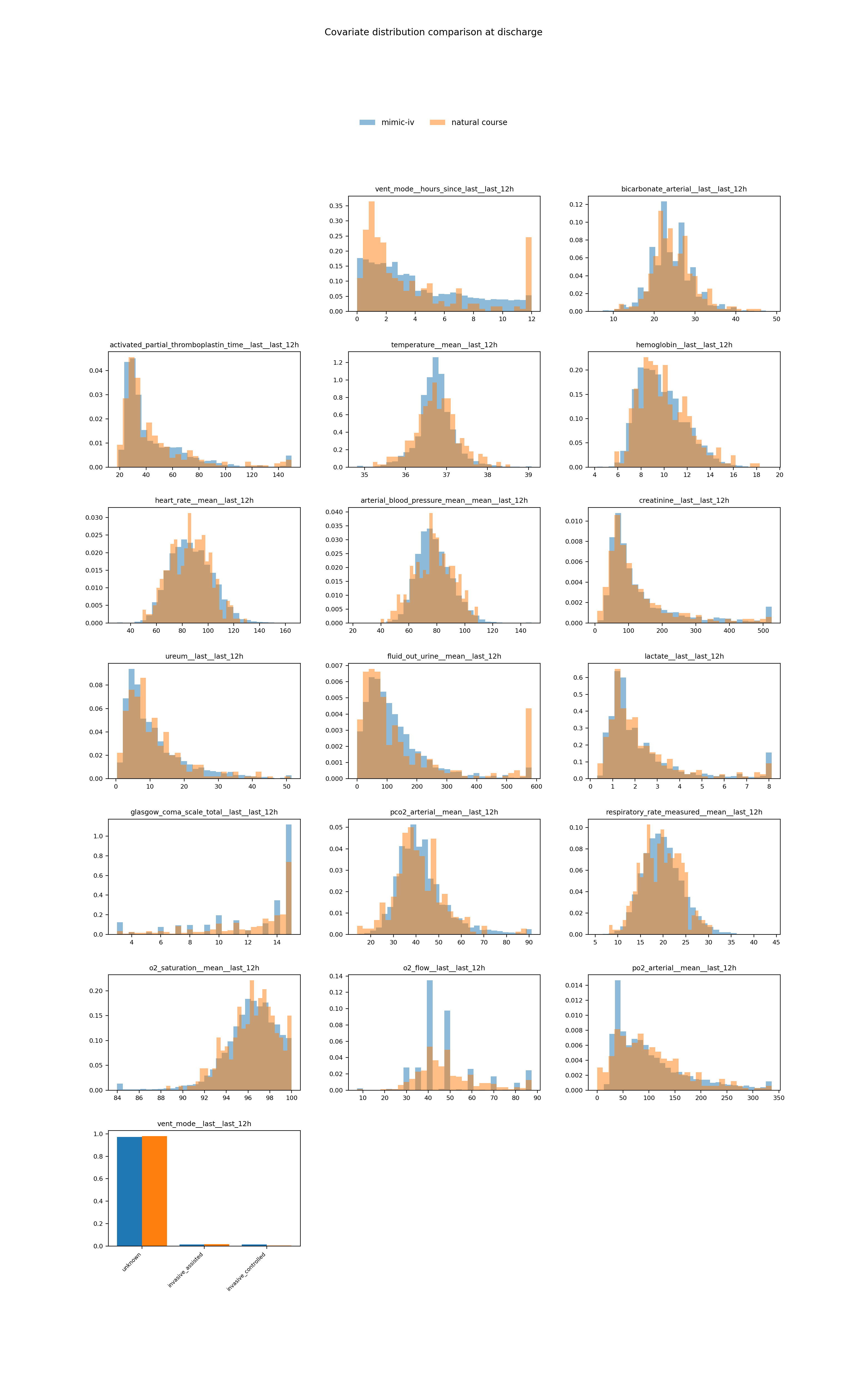}
    \caption{Covariate distribution comparison under the natural course and MIMIC-IV data.
    The close overlap across physiologic variables demonstrates that the natural course simulation reproduces the empirical covariate distributions at discharge.}
    \label{fig:L1_A1}
\end{figure*}

To assess the satisfaction of distributional assumptions, we visually inspected empirical density plots of all time-varying confounders across the 12-hour grid. Figures~\ref{fig:L1_D1} and \ref{fig:L1_A1} display the observed marginal distributions over time and were used to verify consistency with the parametric families specified in the \texttt{pygformula} implementation. Particular attention was paid to skewness, heavy tails, boundary inflation, and temporal drift, ensuring that selected functional forms captured clinically meaningful variability without inducing extrapolation .

\end{document}